\documentclass[preprint,12pt]{elsarticle}
\usepackage{graphicx}
\usepackage{amsmath}
\usepackage{hyperref}
\usepackage{amssymb}
\usepackage{latexsym}
\usepackage{algorithm} 
\usepackage{indentfirst}
\usepackage{wrapfig,lipsum}
\usepackage{epsfig,float}
\usepackage{mathrsfs}
\usepackage{algorithm}
\usepackage{subcaption}
\usepackage{algpseudocode}

\usepackage{graphicx}
  \baselineskip 8 mm
\journal{Online Social Networks and Media}

\begin{document}

\begin{frontmatter}



\title{Identifying Influential Nodes in Weighted Networks using k-shell based HookeRank Algorithm}


\author[add1]{Nipun Aggarwal\thanks{1} }
\ead{nipundtu@gmail.com}

\author[add1]{Sanjay Kumar\thanks{1} }

\address[add1]{Department of Computer Science and Engineering, Delhi Technological University, Main Bawana Road, New Delhi-110042, India}
\begin{abstract}
Finding influential spreaders is a crucial task in the field of network analysis because of numerous theoretical and practical importance. These nodes play vital roles in the information diffusion process, like viral marketing. Many real-life networks are weighted networks, but relatively less work has been done for finding influential nodes in the case of weighted networks as compared to unweighted networks.  In this paper, we propose $k$-shell based HookeRank (KSHR) algorithm to identify spreaders in weighted networks. First, we propose  weighted k-shell centrality of node ($u$) by using k-shell value of $u$, k-shell value of its neighbors ($v$), and edge weight ($w_{uv}$) between  them.  We model edges present in the network as springs and edge weights as spring constants. Based on the notion of Hooke's law of elasticity, we assume a force equal to weighted k-shell value acts on each node. In this arrangement, we formulate KSHR centrality of each node using associated weighted k-shell value and the equivalent edge weight by taking care of series and parallel combination of edges up to 3-hop neighbors from the source node.  The proposed algorithm finds influential nodes that can spread the information to the maximum number of nodes in the network.  We compare our proposed algorithm with popular existing algorithms and observe that it outperforms them on many real-life and synthetic networks suing  Susceptible-Infected-Recovered (SIR) information diffusion model.

\end{abstract}
\begin{keyword}
Complex networks \sep Centrality measure \sep Epidemic Spreading \sep Influential spreaders \sep Social networks \sep Weighted networks


\end{keyword}

\end{frontmatter}


\section{Introduction}

In our day to day life, we come across numerous complex networks like communication networks, social networks, biological networks, and the world wide web network \cite{complex1}.  Such networks consist of a large number of nodes with non-trivial properties leading to a variety of research problems \cite{complex2}. Complex networks can be modeled as graphs $G=(V,E)$  where $V$ corresponds to set of nodes, $E$ represents edges  between nodes.  
Many complex networks are weighted networks where edges are associated with some weight, where weight refers to the duration or frequency of communication, collaboration, friendship, or trade between two entities \cite{Newman}. For example, in social networks, the weight of an edge can be a function of time, affection, and exchange of services between two persons \cite{Granovetter}.  In other weighted networks,  the weight often expresses the potential or capability of the tie \cite{Opsahl}. In airport networks, the edge weight between the two cities can refer to the number of direct flight services between them. Usually, information or diseases are more likely to be transferred between the nodes having high or frequent interactions.


Finding influential spreaders is one of the prominent research problem in the field of complex networks analysis \cite{influential}. Influential nodes play crucial roles in many spreading phenomena like viral marketing, influence maximization \cite{influence1}, rumor control \cite{rumor}, epidemic spreading \cite{spreading}.  The action of influential spreaders can lead to maximizing the information coverage in the advertisement of the products and viral marketing, whereas minimizing the damage in case of rumor or epidemic spreading in the network. Influence maximization \cite{influence2} aims to choose a constant number ($k$) of influential nodes as seed nodes such that information originating from them can reach to the maximum number of nodes in the network due to cascade trigger through word-of-mouth strategy. The influence maximization activity comprises of two phase: identifying the seed spreaders, and information diffusion phase. Researchers have effectively applied various epidemic spreading methods to model information diffusion in networks to assess influence spread arises due to trigger from the selected seed nodes \cite{spreading1}. Identifying influential nodes problem on unweighted networks has been extensively studied in recent years. However, it is relatively less explored for the weighted networks. The majority of such methods on weighted systems are just extension of existing methods on unweighted methods by bringing edge weight into the picture. In this paper, we propose a k-shell decomposition and Hooke's law of elasticity based algorithm named as KSHR  for locating the influential spreaders in the weighted network. Our algorithm considers the influence of nodes in a setting where edges are demonstrated as springs and edge-weights are displayed as flexibility coefficients. We model edge weights as spring constants. The edges present in the system are displayed as springs, which are associated in arrangement and equal. They lengthen by a separation under the influence of an expected steady power adhering to Hooke's law of flexibility, and this is the identical spread separation between nodes in the system. 

The contributions of our work are as follows:
\begin{enumerate}
   
    \item We introduce a new algorithm for weighted k-shell decomposition and use the weighted k-shell value of a node as stretch value for the edges connected to that node in a setting where the weighted graph is modeled as springs.
     \item Based on the notion of weighted k-shell centrality and Hooke’s law of elasticity in the case of complex weighted networks, we propose the KSHR algorithm to identify influential nodes.
    \item The experimental results on real-life and synthetic network datasets based on various performance parameters reveal that the proposed algorithm performs reasonably well compared to other existing popular methods. 
\end{enumerate}
The rest of the paper is organized as follows: Section \ref{related} presents the related weighted centrality,  related applications of modeling the edge of the graph as the spring, and information diffusion model.  Section \ref{datasets} discusses the performance matrices and data-sets used in this paper. In Section \ref{method}, we describe the proposed algorithm, time complexity of the algorithm, and its simulation on a toy network.   Section \ref{results} discusses the experimental results on each performance parameter for various datasets. Finally,  in Section \ref{conclusion}, we conclude our study.

\section{Preliminary}
\label{related}
\subsection{Node Centrality}
The underlying models in the field of impact maximization have significantly been advancements in the field of unweighted networks where all edges are similarly significant. In real networks, 
these edges are related with weights that should be thought of while calculating the quality of these nodes during a pattern of information diffusion. At the point when we think about these parts of topology, we can accumulate insights into what is generally useful for the expansion of information propagation. The early advances in weighted networks were through centralities like DegreeRank utilized for unweighted networks, by extra weighing of these edges to accomplish a weighted DegreeRank \cite{wdegree}. In a similar way, different centralities taken from unweighted networks, advancing into a technique for weighted graphs through numerical changes adjusting the weighted calculations. Betweenness centrality thinks about the most limited way of a node in an unweighted graph, and it was stretched out for weighted rendition giving the weighted-betweenness centrality \cite{between,wbetween}. In view of the thought of the voting method, analysts have proposed influence calculation in unweighted just as weighted networks where the nodes getting the maximum votes in each round as spreader nodes \cite{vote,wvote,ncvote}. The $h$ index is a proportion of the impact of analysts dependent on the quantity of references got, and by expanding edge weight, Yu et al. proposed a weighted $h$-index centrality \cite{hindex}. Weighted-eigenvector centrality is appropriate in a weighted network and depends on the way that a node is significant if its neighbors are likewise celebrated and finds the centrality for a node as an element of the centrality of its neighbors \cite{weigen}. 

\subsection{Spring-based edges} 

Eades \cite{eades} proposed to display the edges of the network as springs to draw graphs by limiting expected potential energy. This technique was later refined by Fruchterman et al. \cite{fruch}, where they model nodes as electrical charges and edges as associating springs. The electrical charges cause these nodes to repulse one another. One of the most well known calculations for drawing graphs is Kamada and Kawai's strategy, which models the edges of the graph as springs acting observing Hooke's Law \cite{kamada,hooke}. The strategy increases the length of the spring between any two nodes by limiting a worldwide cost variable. We contend the relevance of the spring-based model to quantify the centrality of the nodes and to discover persuasive spreaders. 

\subsection{Weighted K shell} 

The first K-shell was applicable to just unweighted graphs \cite{ks}. It is a classical calculation that has been utilized to discover influential nodes. The expected augmentation of the calculation on weighted networks was performed by consideration of weights on biological networks by Eidsaa \cite{eidsaa}. The calculation was additionally stretched out based on interaction and a hybrid k-core methodology was created \cite{al}. These centralities were improved by taking neighboring k-shells and figuring the consolidated result by Maji and Wei \cite{maji, wei}. 
We propose a weighted K-shells method that performs better in our context.

\subsection{Information Diffusion Model} 

SIR Model: In this paper, we employ the popular stochastic susceptible-infected-recovered (SIR) model as the information propagation model \cite{SIR}. The SIR model partitions nodes of the network into three classes: susceptible ($S$), infected ($I$), and recovered ($R$). Susceptible are those nodes that are supposed to receive the information from their infected neighbor. In this model, the selected seed nodes are initially in an infected state, and all other nodes are in a susceptible state. In each subsequent iteration, the infected nodes influence their susceptible neighbors with a probability of \(\beta\). Infected nodes at the following timestamp enter the recovered stage with a probability of \(\gamma\). Once a node is in recovered state, it can't infect its neighbors further. Hence, according to this model, the amount of influence spread caused by the selected seed nodes can be estimated by counting the number of nodes who got infected and then later recovered in the subsequent stage of the SIR simulation.

\section{Datasets and Performance Metrics} 

\label{datasets} 
\subsection{Performance Metrics}
In this paper, we adopt a variety of evaluation matrices to reasonably analyze the performance of the proposed algorithm and other existing popular algorithms. The following evaluation matrices are included in our study:
\begin{enumerate}
    \item [(i)] Influence spread or number of active nodes: It counts the total number of nodes that became active or infected at the end of the simulation of the information diffusion model, which caused due to the cascade effect triggered by the initial spreaders. In the  SIR model of information dissemination, the selected seed nodes are initially infected nodes. They try to infect their susceptible neighbors with the rate $\beta$. This process continues till the state of nodes keeps on changing, and at the end, the total number of nodes during the diffusion process which converts from susceptible state to infected state and then to recovered state are counted to measure the influence spread. It is one of the most important parameters to judge the effectiveness of an influence maximization algorithm as it directly measures the overall spread of the information in the network originating from the selected spreaders. 
\item [(ii)] Kendall tau correlation ($\tau$) :-
Kendall tau correlation  \cite{kendall} measures the ordinal association between two measured quantities in statistics. It is utilized to determine the precise spreading influence of the selected spreader nodes.  In our case, the Kendall tau correlation is used to find the correlation between the rank list ($R_1$)  generated by various algorithms of the influence maximization with the natural ranking list ($R_2$) generated by the SIR Model.  For a pair of seed nodes, if $x_i$ $>$ $y_j$ in the list $R_1$, and $x_i$ $>$ $y_j>$  in actual rank list $R_2$  then such pair is known as concordant pair otherwise it is called discordant pair. The value of Kendall tau correlation lies in the range [-1, 1]. Kendall tau correlation ($\tau$) between two ranked list $R_1$ and $R_2$ is computed using the following formula:

\begin{equation}
\label{eqn11}
\tau(R_1 , R_2) = \dfrac{n_{con} - n_{dis}}{ \dfrac{1}{2} n(n-1) }
\end{equation}
where $n_{con}$ is the number of concordant pairs, $n_{dis}$ is the number of discordant pairs and $n$ is total number of nodes in both lists $L_1$ and $L_2$.

\item [(iii)] Average distance between spreaders ($L_s$): 	This metric computes the average shortest path distance between the selected spreaders. 
The high value of the average distance between spreaders implies that the spreaders are chosen from the diverse locations in the network and therefore, can propagate information to a significant portion of the system. 
The following equation computes the average distance between spreaders ($L_s$):
\begin{equation}
    L_s=\frac{1}{\frac{|S|(|S|-1)}{2}}\sum_{u,v \in S\ u\neq v} d_{uv}
\end{equation}
where $S$ represents the set of nodes and $d_{uv}$ represents the shortest path distance between a pair of seed nodes $u$ and $v$ in $S$.
\end{enumerate}

\subsection{Datasets used}
\begin{table}[H]
\caption{Real world data-sets for simulation}\label{tab1}
\begin{tabular}{|l|l|p{7cm}|l|l|}
\hline
S.No. &Dataset Name &  Description & \#Nodes & \#Edges\\

\hline
1 &Newman-CondMat \cite{cond} & This is the co-authorship network of reprints posted to the Condensed Matter section of the arXiv E-Print Archive. & 16264&47594\\
\hline
2 &US-Airports \cite{airports}&The data is downweighted from the Bureau of Transportation Statistics (BTS) Transtats site (Table T-100; id 292) with the following filters: Geography=all; Year=2010; Months=all; and columns: Passengers, Origin, Destination & 2939 & 15677\\
\hline
3& Bitcoin \cite{bt} & The datasets contains the details of the members of Bitcoin OTC rate other members in a scale of -10 (total distrust) to +10 (total trust) in steps of 1. & 5881 & 35592\\
\hline
4 &Facebook-like  \cite{fblike} & This is a weighted dataset originates from an online community for students at the University of California, Irvine & 1899 & 13838\\
\hline
5 & p2p-Gnutella \cite{p2p} & A sequence of snapshots of the Gnutella peer-to-peer file sharing network from August 2002. & 6301 & 20777\\
\hline

6 & Barabasi-2000 \cite{bara} & An undirected randomly weighted network based on growth and preferential attachment in scale-free networks. & 2000 & 7984\\
\hline
7 & Barabasi-9000 \cite{bara}  & An undirected randomly weighted network based on growth and preferential attachment in scale-free networks. & 9000 & 35984\\
\hline
\end{tabular}
\end{table}

\section{Methodologies}
\label{method}
Hooke's law is stated as follows:
\begin{equation}
\label{Hooke}
    F = k.x
\end{equation}
In this section, we present the depiction of the edges present in the network as springs, which are associated in series and parallel, and present the proposed KSHR algorithm. In a weighted network, weights commonly imply that the higher the weight, the stronger is the association. The same is valid for our KSHR centrality, since we know from Hooke's law that more is the constant of the spring($k$) and less is the force ($F$), lower is the spring stretched ($x$) as in eq. \ref{Hooke}. When we visualize the edges of the graphs as springs, where their edge-weight takes the role of spring constant $k$, the separation ($x$) is the parameter to be minimised between the nodes for the influence maximization. By calculating the separation ($x$), we can accommodate the various paths that exist between any pair of nodes. We describe the definition of spring terminology below.\\

\textbf{Parallel: } When Springs are put in parallel, they end up as a joint spring with the complete elasticity of another spring of a spring stretched that can be displayed utilizing the fact that the spring is certainly much stiffer.\\ 

\textbf{Series: } It is conceivable to include the commitments of the springs in series. The Springs in series make a progressively increasing spring that will in general stretch more than the springs that are connected.\\

\subsection{Distance Calculation}
When two springs of different spring constants, \(k_1\) and \(k_2\) respectively are placed in
series with each other, we get:
\begin{equation}\label{series}
    \frac{1}{k_{eq}} = \frac{1}{k_1} + \frac{1}{k_2}
\end{equation}

When two springs of different spring constants, \(k_1\) and \(k_2\) respectively are placed in
parallel with each other, we get:

\begin{equation}\label{parallel}
    k_{eq} = k_1 + k_2
\end{equation}

 This means in an actual network is that springs in series are stiffer if the strength of ties in the individual connections is strong. It also implies that more connections from one node to another, add up a single connection, as seen in case of parallel connections. Now the equivalent distance between any of these nodes, under a constant force is given as:

\begin{equation} \label{distance}
    x = \frac{f}{k}
\end{equation}
where $f = weighted-k\_shell value$ and $k =$ weight of the edge between node $i$ and $j$
When we model edges present in the graph as springs, which are associated in series and parallel. 
\begin{figure}[h!!] 

\centering 

\includegraphics[width=7cm]{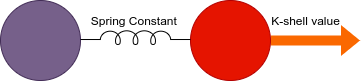} 

\caption{Calculating the heuristic separation between two nodes dependent on weights of the edges, being displayed utilizing springs and assessing to a solitary spring, adhering to Hooke's Law of Elasticity} 

\label{eq} 

\end{figure} 

As shown in Fig. \ref{eq}, an amount of force acts on each node and hence causes the springs to be stretched. For our method, we consider a constant force acting on each node, which is given by its weighted k-shell value. Thus, the force on every node is given as \(F_0 = wkshell[node]\), which stretches the edges connected to it. We propose a novel weighted k-shell value in this paper. 

Our model assumes the node($i$) is associated with direct neighbor node ($j$) with a spring constant $k$ equal to edge weight of $w_{ij}$. This gives us \(k = w_{ij}\) as the spring constant of direct neighbors. 
By exploiting series and parallel combination of springs, for a given source node we can consider its indirect neighbors as well for the information spread. The value of spring constant $k_{eq}$ for indirect neighbors can be computed using the formula for series and parallel combinations (Eq. \ref{series}, and \ref{parallel}). Now, using the value of $k_{eq}$, we compute equivalent spring stretch distances ($x$) for direct and indirect neighbors. In our algorithm, we consider the indirect neighbors up to 3-hops by utilizing a Breadth first traversal (BFS) for a given source node ($i$). Let us consider the node ($i$) with which our BFS starts and $j$ as one of its indirect neighbors. We calculate the spring constant ($k_{eq}$) between nodes $i$ and $j$. The stretch distance ($x$) between nodes $i$ and $j$, is given by Eq. \ref{distance}. Our overall objective here is to minimise the value of stretch distance ($x$) for the neighbor because a less stretch distance shall mean an increased spreading potential of seed node $i$ implying that information spread by $i$ can easily reach $j$. Thus, restating the same as follows:
\begin{equation} \label{maximize}
    \frac{1}{x} = \frac{k}{f}
\end{equation}
Assuming $f = wkshell[node]$, we can easily see that the full measure of distance in this network is relative. We need to maximise the value of $1/x$ so that the spreading can happen and we name the average of this value as $kshr\_value$ of the given node $i$ when connected to all nodes $j$ such that $j \epsilon N^3_i$, where $N^3_i$ signifies the neighbors upto 3 hops.
\begin{equation} \label{val}
    kshr\_value[i] = \frac{(\sum_{j \epsilon N^3_i} w_{ij})} { wkshell[i]}
\end{equation}



\subsection{Calculation of K-Shell Value}

\begin{equation}
    \label{wk}
    wk_{i} = k\_shell[i] + \sum_{j \epsilon N_i} \sqrt{w_{ij}*k\_shell[j]}
\end{equation}
\begin{algorithm}[H] 
\caption{Weighted K-shell Algorithm}
\label{alg:kshell}
\begin{algorithmic}[1]
\Require{$Graph$} 
\Statex
\Function{WKShells}{$G$}
  \State {$nodes$ $\gets$ {$list(G.nodes())$}}
    \State {$wkshell$ $\gets$ {$\{\}$}}
    \For{$node$ in $nodes$}                    
        \State {$wkshell[node]$ $\gets$ $kshell[node]$}
		\State {$neighbors \gets list(G.neighbors(node))$}
		\State {$s \gets 0$}
		\For{$neighbor$ in $neighbors$}   
			\State {$wij \gets G[node][neighbor][weight]$}
			\State {$s += \sqrt{(wij*kshell[neighbor])}$}
		\EndFor
		\State {$wkshell[node] += s$}
    \EndFor
    \State \Return {$wkshell$}
\EndFunction
\end{algorithmic}
\end{algorithm}
\subsection{Proposed algorithm}
The explanation of the algorithm is as follows:
\begin{enumerate}
    \item We create a $kshr\_value$ dictionary and initialize for all nodes as in Step 1.
    \item Since we wish to minimize $x$, the stretch distance of the string. We instead maximize $1/x = k/f$ as in Eq. \ref{maximize}. Hence, we create a dictionary named as $spring\_constant$ of constants from Step 4 and 5 to calculate $k$.
    \item Our BFS happens for 3 hops from the start node in the complete loop from 7 to 13, here $N^3_i$ denotes all neighbors upto 3 hops.
    \item The BFS takes the nodes that have not been visited and initializes them to the value of the parent and combines them in series in Step 9 using equation \ref{series}.
    \item All the nodes that occur again in the BFS traversal are assumed to be in a parallel connections and add up to the spring constant as in Step 11 as in eq. \ref{parallel}.
    \item We calculate the inverse distance for the node by finding the equivalent spring constant ($k_{eq}$ between the neighbor and the current node and dividing it by the node's weighted K-shell value as in step 16 given by Eq. \ref{maximize}
    \item The KSHR Value is calculated based on the addition of these individual distances given by Eq. \ref{val} 
    \item As, the objective of influence maximization is to select top $c$ nodes, where $c$ is a constant. Here,  the top $c$ nodes are the nodes having the maximum KSHR score values in the ranking, and such nodes can be chosen as the influential spreaders.
    \item We sort the KSHR value dictionary and return it so that top nodes can be chosen as influential spreaders.
\end{enumerate}
\subsection{Time Complexity Analysis}
The time complexity of this algorithm is calculated as follows. We traverse on each node of the graph and venture to 3 hops of its neighbors. The steps from line 15-34 give us a time complexity of \(O(k^3.n)\). If we assume hat the diameter of the network is fairly large, the average value of \(k = n^{1/d}\). Thus giving a loose upper bound of \(O(n^{\frac{d+3}{d}}\). In sparse graphs, $k$ is a constant with much lower magnitude and the average complexity is of the order of \(O(n)\)
\begin{algorithm}[H] 
\caption{KSHR Algorithm}
\label{alg:kshr}
\begin{algorithmic}[1]
\Function{KSHR}{$G$}
    \State {$kshr\_value$ $\gets$ {$\{\}$}}
    \For{each $node$ in $graph$}                    
        \State {$kshr\_value[node]$ $\gets$ $0$}
        \State {$spring\_constant$ $\gets$ {$\{\}$}}
        \State {$N^3_i \gets THREE\_HOP\_NEIGHBORS(G, node)$}
	    \For{$neighbor$ in $N^3_i$ perform BFS}
	        \If{$neighbor$ in series}
	            \State {use Eq \ref{series}} for series combination
	        \Else
	            \State {use Eq \ref{parallel}} for parallel combination
	        \EndIf
	    \EndFor
	    \State wkshell = $WKSHELLS(G)$
		\For{$neighbor$ in $N^3_i$}
		    \State $kshr\_value[node]+=spring\_constant[neighbor]/wkshell[node]$
		\EndFor
    \EndFor
    \State sort($kshr\_values$)
    \State \Return {$kshr\_value$}
\EndFunction
\end{algorithmic}
\end{algorithm}

\section{Results and Analysis}
 \label{results}

We perform the experiment of the proposed KSHR method along with the contemporary centrality measures like weighted- degree, weighted betweenness centrality, weighted eigenvector centrality, and weighted voteRank. The investigation has been performed on a toy network and three real-world networks of different nature, application, and size that are listed in Tab \ref{tab1}. We use the SIR model to compute the final infected scale, $f(t\textsubscript{c})$, as a function of spreaders fraction and final infected scale in terms of increasing timestamps. The results were averaged over SIR $100$ simulations. For simplicity and to maintain consistency in the analysis for all data-sets, we chose infection rate ($ \beta $ ) as $0.01$, meaning that when a node is infected, then it can infect 1$\%$  of its neighbors randomly.

\subsection{Simulation of the proposed algorithm on a toy network}
\begin{figure}[H]
\centering

        \includegraphics[width=4in,height=2in]{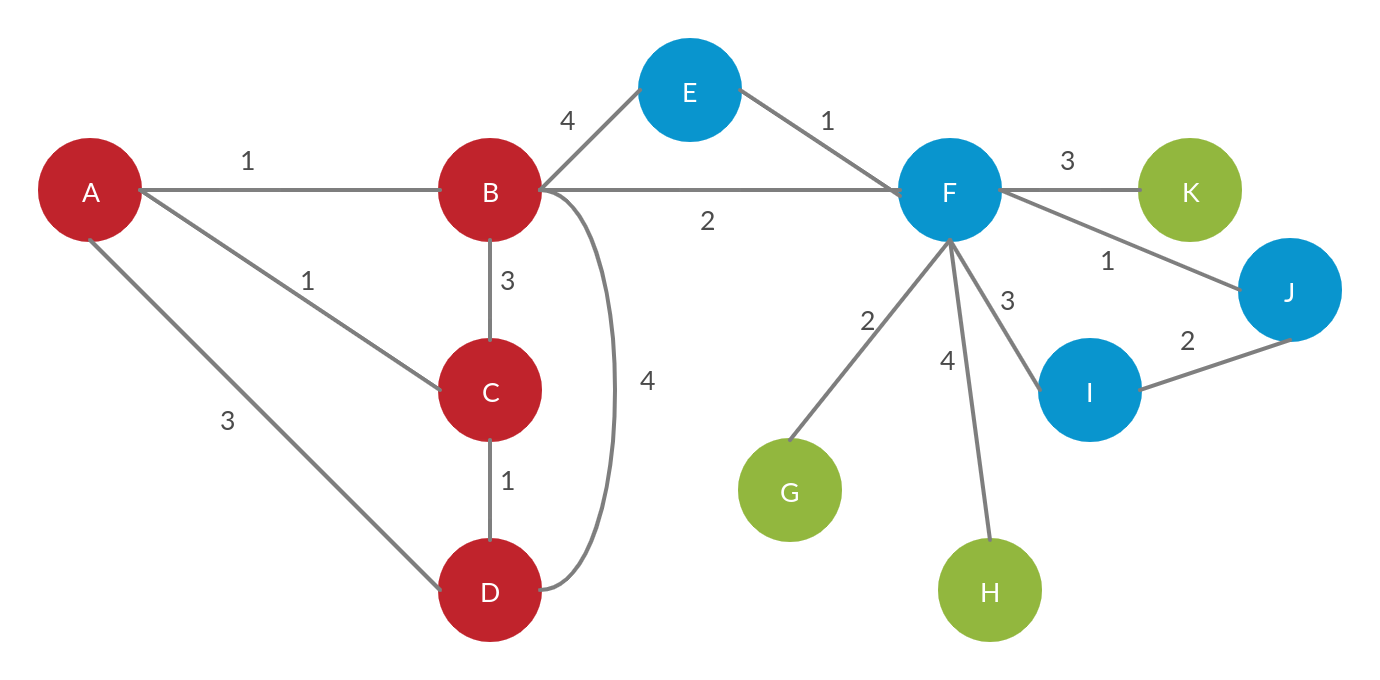}
        \caption{Weighted toy network with 11 nodes to simulate the proposed KSHR centrality}\label{col}
\end{figure}

In this section, we simulate the working of the proposed KSHR method using a toy network, as depicted in Fig. \ref{col}. The toy network that we use is randomly weighted and has 11 nodes with different edge weights. The force acting on each node is assumed to be equal to the weighted-k-shell value of the node. The traditional K-shell value of the nodes is indicated using colors in Fig. \ref{col}. The red colored nodes have K-shell value 3, the blue colored nodes have value 2, and the green colored nodes have a K-shell value 1. The traditional K-shell value will help us calculate the weighted K-shell value as in Eq. \ref{wk}.

\begin{figure}[H]
\centering

        \includegraphics[width=2.5in,height=1in]{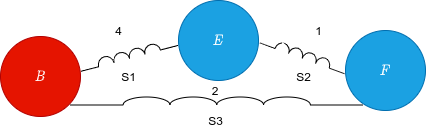}
        \caption{Spring S1 and S2 are in series and the resulting spring is in parallel with S3}\label{sp}
\end{figure}

 Nodes are connected using weighted-edges represented as springs, based on our algorithm. We aim to find the equivalent spring between all direct and indirect neighbors of a given node using series and parallel spring combination. This relation between series and parallel is illustrated through the example of nodes $B$ and $F$ in this case. As shown in Fig. \ref{sp}, the spring between $B$ and $E$ is in series with the spring from $E$ to $F$. The resulting value of resulting spring $B-E-F$ is found using eq. \ref{series}. This resulting spring $B-E-F$ is in parallel to the spring between $B$ and $F$ and the net spring between $B$ and $F$ is given by eq. \ref{parallel}. 

Let us consider the steps of the algorithm for a sample node $B$ to find the KSHR-value of $B$ and understand the working of KSHR centrality. Now, we begin processing the algorithm by taking the node $B$ and compute the spring constant for the 1-hop neighbors of $B$, namely $A$, $C$, $E$, $D$, and $F$ using the series connection of the spring. After the 1-hop neighbors are covered and we get the 2-hop neighbors by combination series in parallel similar to the example of $B$, $E$, and $F$ above. A similar calculation is performed taking all nodes in the network as the start and performing the computation using series and parallel combination of the direct and indirect weighted edges to calculate the KSHR-value given by eq. \ref{val}. Table \ref{toyres} summarises the computed KSHR-values and $F$ is determined to be the top node for the toy-network.

\begin{table}[H]

\caption{Results on the toy network}\label{toyres}
\begin{tabular}{|l|p{5cm}|l|l|}
\hline
Node & K-Shell & WK-Shell & KSHR-Score\\
\hline
A & 3 & 14.660254037844386 & 0.40297842100090797\\
B & 3 & 40.88098600800384 & 1.0447729815375513\\
C & 3 & 15.928203230275507 & 0.3469518067328419\\
D & 3 & 20.660254037844386 & 0.5588701059990411\\
E & 2 & 54.61172454625088 & 0.3578790830315174\\
F & 2 & 10.342416792648603 & 3.356840317431281\\
G & 1 & 3.0 & 0.36995550589091647\\
H & 1 & 3.8284271247461903 & 0.36995550589091647\\
I & 2 & 8.898979485566356 & 0.6287575668893135\\
J & 2 & 6.82842712474619 & 0.46272587178047\\
K & 1 & 3.449489742783178 & 0.4897801816713282\\
\hline
\end{tabular}
\end{table}





 \subsection{Simulation of the proposed algorithm on real-life networks}

\begin{figure}[H]
    \centering
        \includegraphics[width=120mm]{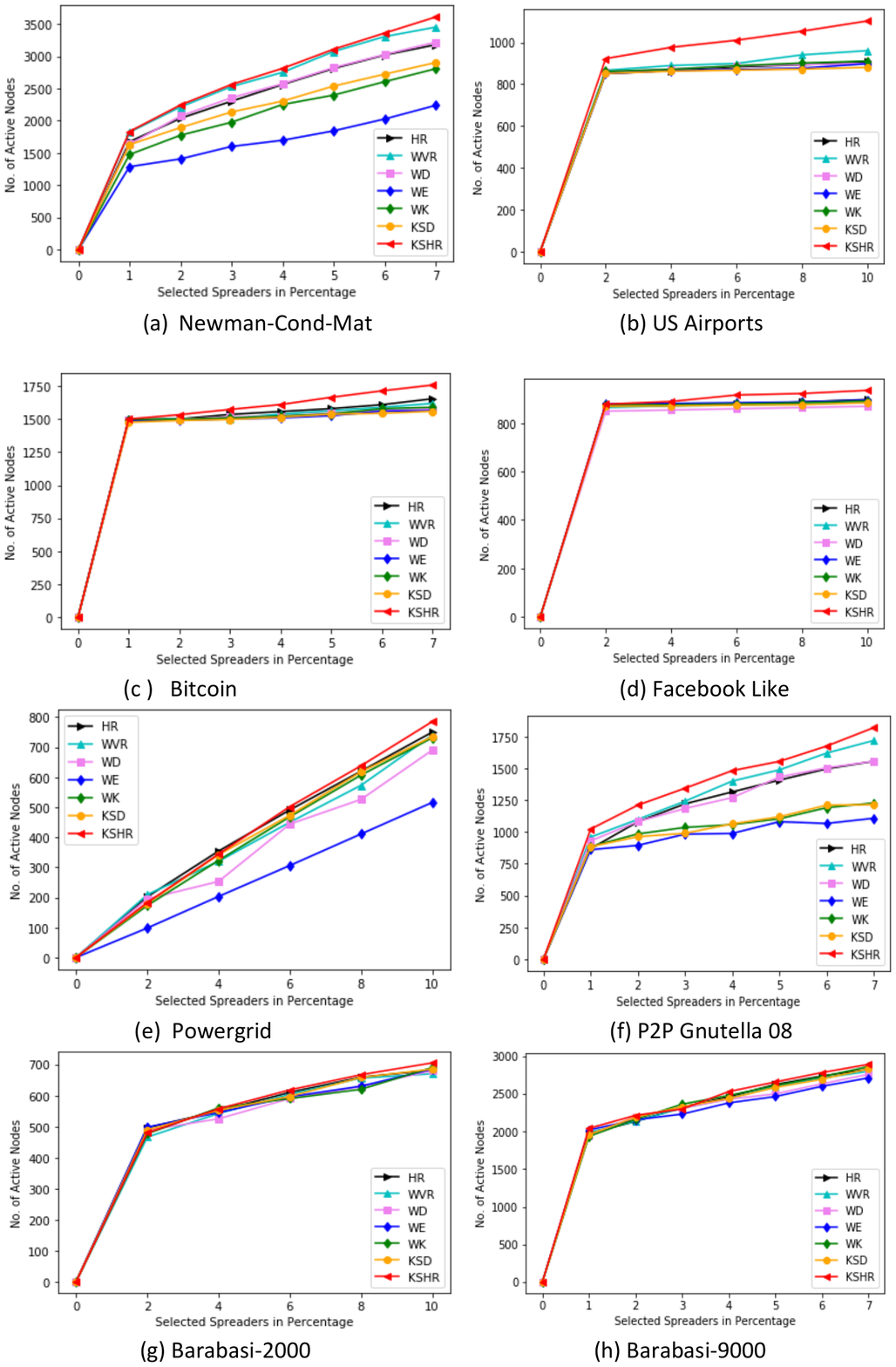}
    \caption{The infection scale with respect to the percentage of spreaders on Facebook-like weighted network with $ \beta $  = 0.01.}
    \label{fig8}
\end{figure}

\begin{figure}[H]
    \centering
        \includegraphics[width=120mm]{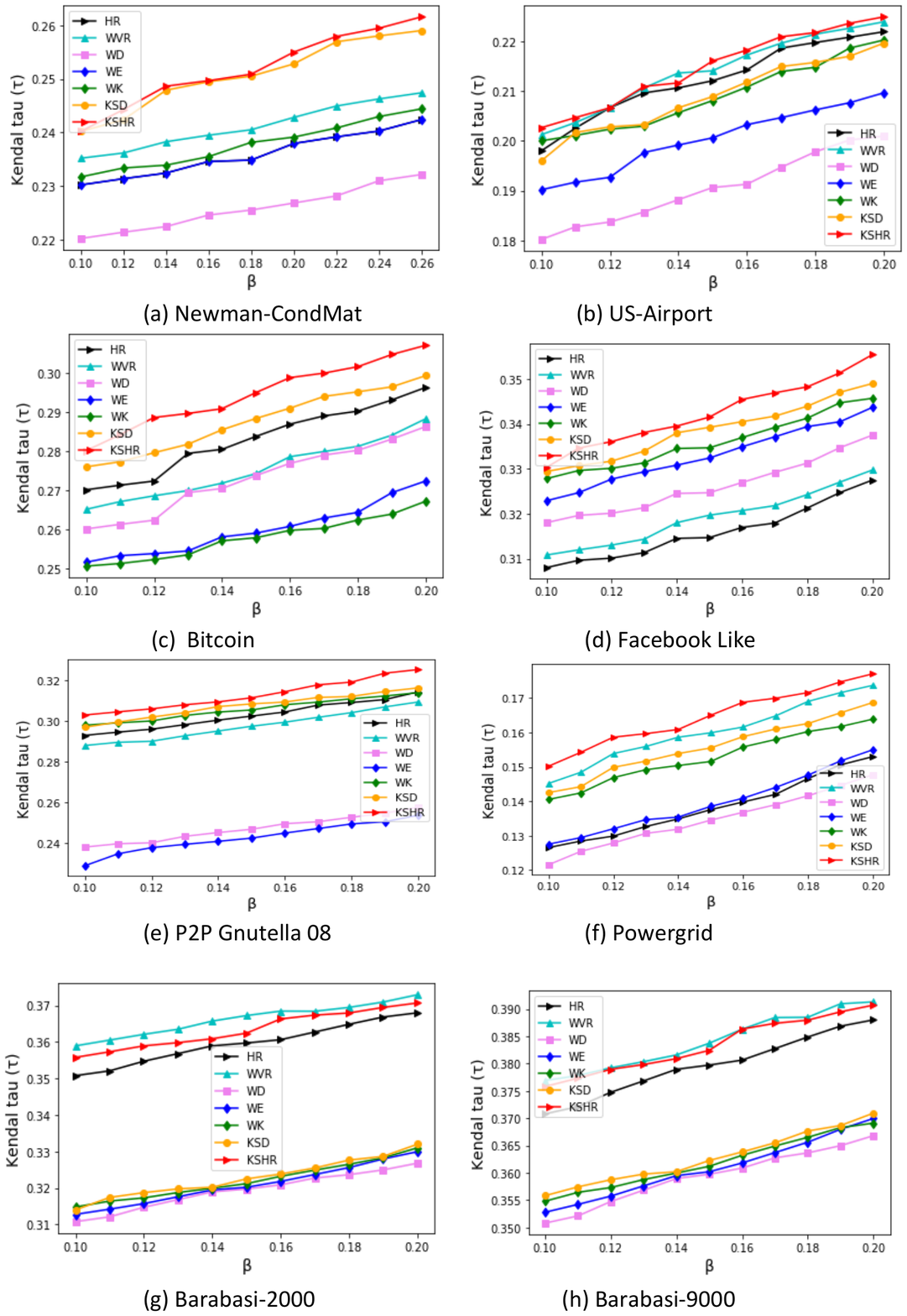}
    \caption{tau vs beta}
    \label{fig9}
\end{figure}

 Fig. \ref{fig8}, and Fig. \ref{fig9} depicts the final infection scale ($f(t_c)$) with respect to the percentage of spreaders for three real-life data-sets with infection rate ($\beta$) as $0.01$. We consider the percentage of influential spreaders as the seed nodes in the range of $2\%$, $4\%$, $6\%$, $8\%$, and $10\%$ to plot the final infection scale.
 In  Fig. \ref{fig7}, note that the number of nodes affected by the infection is maximum for HookeRank on the US-Airports Network for most percentages of spreaders. In Fig. \ref{fig8}, HookeRank greatly exceeds the performance of other algorithms towards increasing the count of the spreader fraction. In the weighted PowerGrid data, shown in Fig. \ref{fig9}, HookeRank performs better than most other algorithms from an early stage. In the weighted PowerGrid data, shown in Fig. \ref{fig10}, the increase in the number of spreaders results in WVoteRank becoming marginally close to HookeRank, but our algorithm still performs better than all other algorithms in the simulation.



\noindent 


\noindent 

Fig. \ref{fig10} presents the final infection scale ($f(t_c)$) with respect to the increasing timestamps with infection rate ($\beta$) as $0.01$ and top $7\%$ influential as seed nodes on US PowerGrid network.
Fig. \ref{fig11} shows the final infection scale ($f(t_c)$) with respect to the increasing timestamps with infection rate ($\beta$) as $0.01$ and top $5\%$ influential as seed nodes on US PowerGrid network. Fig. \ref{fig12} displays the final infection scale ($f(t_c)$) with respect to the increasing timestamps with infection rate ($\beta$) as $0.01$ and top $5\%$ influential as seed nodes on Facebook-like weighted network.

\noindent 

\noindent 



\begin{table}[h]
	\centering
	\begin{tabular}{ |l|c|c|c|c|c|c|c|c|}
		\hline
		Dataset &$L_s(HR)$&$L_s(WVR)$&$L_s(WD)$ &$L_s(WE)$&$L_s(WK)$&$L_s(KSD)$&$L_s(KSHR)$\\
		\hline
		Newman-CondMat   & 2.80&	2.925&	2.80&	1.18&	1.95&	2.45& \textbf{5.86}\\
		US Airports       &	1.91 &	1.82 &	1.81 &	1.84&	1.72&	1.75 & \textbf{4.02}\\
		Bitcoin  & 25.02&	24.94&	24.31 &	22.41&	23.20 & 24.02 & \textbf{34.02}\\
		Facebook Like & 2.01 &1.88 &2.02 &\textbf{2.29} &1.81 &1.81 &2.05\\                  
		Powergrid &	13.45&	14.71 &	13.45&	1.92&	5.41&	\textbf{16.04} & 15.17\\
		P2P Gnutella 08  &	5.46&	4.77&	5.18&	5.39&	4.59&	4.59 & \textbf{8.34}\\
		Barabasi-2000   &	25.62&	25.62&	25.62&	\textbf{30.8}&	27.10 &	27.10& 26.34\\
		Barabasi-9000   &23.62& 22.73& 22.73& 24.11&22.73& 22.73& \textbf{26.62}
\\        
		\hline
	\end{tabular}
	\caption{ Average Shortest Path Length ($L_s$) values between selected spreaders  where
	      $HR$, $WVR$, $WD$, $WE$, $WK$, $KSD$, and $KHSR$ represents HookeRank, weighted VoteRank, weigted degree,  weighted Eigenvector, weighted $k$-shell, weighted $k$-shell with, and proposed weighted $k$-shell based HookeRank   algorithm respectively. }
	\label{tab4}
\end{table}

From above results on three real-life networks, it is evident that HookeRank performs better than state-of-the-art methods like weighted-degree centrality, weighted-betweenness centrality, weighted-eigenvector centrality, and weighted-voteRank, and also stretchedly outperforms recent methods like WVoteRank in terms of final infected scale with respect to time $t$ and spreader fraction $p$ on real-world networks as depicted in Tab. \ref{tab0}.

\section{Conclusion}
\label{conclusion}
In this paper, we proposed the KSHR method for finding influential nodes in weighted networks by modeling edges of the network as springs and edge weights as spring constants. Initially, we found a measure of the distance between indirect neighbors through the series and parallel combination of edges, by modeling them as springs. Then we proposed a new method of calculating K-shell score on a weighted network. By finding the KSHR value of the nodes, our method locates the top spreaders in the given real-world network to reach a large number of people in the network to maximize the spread of the information. We performed the simulation of the proposed method along with contemporary methods on six real-life data-sets taking the basis of evaluation as the average distance, Kendall tau and Final Infected scale and concluded that the proposed influence maximization KSHR centrality performs considerably well and is effective in real-life scenarios.
%
%


\begin{thebibliography}{}
\bibitem{complex1}
R. Cohen, S. Havlin,  Complex Networks: Structure, Robustness
and Function. Cambridge University Press (2010).
\bibitem{complex2}
M. E. Newman,  The structure and function of complex networks.
SIAM Review, 45 (2003) 167–256.

\bibitem{Newman}
 M.E. Newman,  Analysis of weighted networks. Physical Review E, 70(5) (2004) p.056131.
 \bibitem{Granovetter}
 Granovetter, 1977. The strength of weak ties. In Social networks (pp. 347-367). Academic Press.
 \bibitem{Opsahl}
 T. Opsahl, V. Colizza, P. Panzarasa,  J. J. Ramasco,  Prominence and control: The weighted rich-club effect. Physical Review Letters 101 (2008) 168702.
 
 
\bibitem{influence1} 
D. Kempe, J. Kleinberg, É. Tardos, Maximizing the spread of influence througha social network, Proceedings of the Ninth ACM SIGKDD International Conference on Knowledge Discovery and Data Mining (2003)137–146.
\bibitem{influential}
F. Bauer, J. T. Lizier, Identifying influential spreaders and efficiently
estimating infection numbers in epidemic models: A walk counting
approach, EPL (Europhysics Letters) 99 (2012) 68007.

\bibitem{zhu} Zhu, Yu-Xiao, et al. “Social Contagions on Weighted Networks.” Physical Review E, American Physical Society, (2017) 
\bibitem{rumor}
B. Doerr, M. Fouz, T. Friedrich, Why rumors spread so quickly in social net-works, Commun. ACM 55(6) (2012) 70–75.
\bibitem{spreading}
R. Pastor-Satorras, A. Vespignani, Epidemic spreading in scale-free networks, Phys. Rev. Lett. 86 (2001) 3200–3203.


\bibitem{spreading1} W. Wang, M. Tang, , H.E. Stanley, L.A. Braunstein, Unification of theoretical approaches for epidemic spreading on complex networks. Reports on Progress in Physics, 80(3) (2017) p.036603.

\bibitem{mobile} Peng, Sancheng, et al. “Social Influence Modeling Using Information Theory in Mobile Social Networks.” Information Sciences, Elsevier, (2016)

\bibitem{Valente4}
Valente, T.W.: Network models of the diffusion of innovations (No. 303.484 V3)(1995)


\bibitem{influence2} Chen, W., Wang, C., Wang, Y.: Scalable influence maximization for prevalent viral marketing in large-scale social networks. In Proceedings of the 16th ACM SIGKDD international conference on Knowledge discovery and data mining (pp. 1029-1038). ACM. (2010).



\bibitem{spreading2} Sun, Y., Liu, C., Zhang, C.X. and Zhang, Z.K.: Epidemic spreading on complex weighted networks. Physics Letters A, 378(7-8), pp.635-640 (2014).


\bibitem{wdegree} Opsahl, T., Agneessens, F. and Skvoretz, J.: Node centrality in weighted networks: Generalizing degree and shortest paths. Social Networks, 32(3), pp.245-251 (2010).
\bibitem{between} Prountzos, D. and Pingali, K.: Betweenness centrality. ACM SIGPLAN Notices, 48(8), p.35.(2013).
\bibitem{wbetween} Wang, H., Hernandez, J., and Van Mieghem, P.: Betweenness centrality in a weighted network (2008).

\bibitem{vote} Zhang, J., Chen, D., Dong, Q. and Zhao, Z.: Identifying a set of influential spreaders in complex networks. Scientific Reports, 6(1) (2016).
\bibitem{wvote} Sun, H., Chen, D., He, J. and Ch’ng, E.: A voting approach to uncover multiple influential spreaders on weighted networks. Physica A: Statistical Mechanics and its Applications, 519, pp.303-312. (2019).
\bibitem{ncvote} Kumar, S. and Panda, B.S.: Identifying influential nodes in Social Networks: Neighborhood Coreness based voting approach. Physica A: Statistical Mechanics and its Applications, p.124215.(2020).

\bibitem{hindex} Yu, S., Gao, L., Wang, Y.F., Gao, G., Zhou, C. and Gao, Z.Y.: Weighted H-index for identifying influential spreaders. arXiv preprint arXiv:1710.05272.(2017).
\bibitem{weigen} Bihari, A., Pandia, M. K.: Eigenvector centrality and its application in research professionals’ relationship network. 2015 International Conference on Futuristic Trends on Computational Analysis and Knowledge Management (ABLAZE).(2015)
\bibitem{eades} Eades, P.,  Lin, X.: Spring algorithms and symmetry. Theoretical Computer Science, 240(2), 379–405. (2000).
\bibitem{fruch} Fruchterman, T.M., Reingold, E.M.: Graph drawing by force‐directed placement. Software: Practice and experience, 21(11), pp.1129-1164.(1991)
\bibitem{kamada} Kamada, T. and Kawai, S.: An algorithm for drawing general undirected graphs. Information processing letters, 31(1), pp.7-15. (1989).

\bibitem{hooke} Slaughter, William S.: The Linearized Theory of Elasticity. Birkhäuser (2001).
\bibitem{ks} Batagelj, V., and M. Zaversnik. “An O(m) Algorithm for Cores Decomposition of Networks.” ArXiv.org, 25 Oct. 2003, arxiv.org/abs/cs/0310049.
\bibitem{eidsaa} Eidsaa, Marius. “Core Decomposition Analysis of Weighted Biological Networks.” NTNU Open, NTNU, 1 Jan. 1970, ntnuopen.ntnu.no/ntnu-xmlui/handle/11250/2399913.

\bibitem{al} Al-garadi, Mohammed Ali, et al. “Identification of Influential Spreaders in Online Social Networks Using Interaction Weighted K-Core Decomposition Method.” Physica A: Statistical Mechanics and Its Applications, North-Holland, 4 Nov. 2016, www.sciencedirect.com/science/article/pii/S0378437116308068.
\bibitem{maji} Maji, G. (2019). Influential Spreaders Identification in Complex Networks with Potential Edge Weight based k-shell Degree Neighborhood Method. Journal of Computational Science, 101055. doi:10.1016/j.jocs.2019.101055 

\bibitem{wei} Wei, Bo, et al. “Weighted k-Shell Decomposition for Complex Networks Based on Potential Edge Weights.” Physica A: Statistical Mechanics and Its Applications, North-Holland, 11 Nov. 2014, www.sciencedirect.com/science/article/pii/S0378437114009637.

\bibitem{SIR}Hethcote, H.W.: The mathematics of infectious diseases. SIAM Review, 42(4), pp.599-653.(2000).

\bibitem{bara} Yook, S. H., Jeong, H., \& Barabási, A. L. (2002). Modeling the Internet's large-scale topology. Proceedings of the National Academy of Sciences, 99(21), 13382-13386.
\bibitem{fblike} T. Opsahl and P. Panzarasa: Clustering in Weighted Networks, Social Networks, Vol. 31, No. 2, May, pp. 155-163. doi:10.1016/j.socnet.2009.02.002 (2009).
\bibitem{cond} Newman, M. E. J., 2001. The structure of scientific collaboration networks. PNAS 98, 404-409.
\bibitem{airports} Watts, D. J., Strogatz, S. H.: Collective dynamics of “small-world” networks. Nature 393, 440-442 (1998).
\bibitem{bt} S. Kumar, F. Spezzano, V.S. Subrahmanian, C. Faloutsos. Edge Weight Prediction in Weighted Signed Networks. IEEE International Conference on Data Mining (ICDM), 2016.
\bibitem{liu} Sun, Y., Liu, C., Zhang, C.-X., \& Zhang, Z.-K. (2014). Epidemic spreading on weighted complex networks. Physics Letters A, 378(7-8), 635–640. doi:10.1016/j.physleta.2014.01.004 
\bibitem{p2p}J. Leskovec, J. Kleinberg and C. Faloutsos. Graph Evolution: Densification and Shrinking Diameters. ACM Transactions on Knowledge Discovery from Data (ACM TKDD), 1(1), 2007.
\end{thebibliography}
\end{document}